\documentclass[a4paper,11pt]{article}

\usepackage{amsmath,amsfonts,amssymb,graphicx,epsfig,multirow}
\usepackage[usenames]{color}
\usepackage{pstricks}
\usepackage{arydshln,cite}
 \graphicspath{{./eps/}}
\numberwithin{equation}{section}
\definecolor{purple}{rgb}{1,0,1}
\definecolor{darkpurple}{rgb}{1,.2,1}
\definecolor{pink}{rgb}{1,.7,.7}

\newcommand{\nc}{\newcommand}
\nc\disp{\displaystyle}
\nc{\fh}{\hat{f}}
\nc{\muh}{\hat{\mu}}
\nc{\nuh}{\hat{\nu}}
\nc{\spos}[2]{\makebox(0,0)[#1]{$\sm{#2}$}}
\nc{\sm}[1]{{\scriptstyle #1}}
\nc{\qbar}{\overline{q}}
\def\Tr{\mathop{\mbox{Tr}}}
\nc{\bib}{\bibitem}
\nc{\al}{\alpha}
\nc{\g}{\gamma}
\nc{\G}{\Gamma}
\nc{\D}{\Delta}
\nc{\eps}{\epsilon}
\nc{\la}{\lambda}
\nc{\La}{\Lambda}
\nc{\var}{\varphi}
\nc{\pa}{\partial}
\nc{\nn}{\nonumber \\ }
\nc{\hf}{\frac{1}{2}}
\nc{\dz}{\frac{dz}{2\pi i}}
\nc{\bin}[2]{\left(\!\!\!\begin{array}{c} {#1}\\ {#2} \end{array}\!\!\!\right)}
\nc{\be}{\begin{equation}}
\nc{\ee}{\end{equation}}
\nc{\bea}{\begin{eqnarray}}
\nc{\eea}{\end{eqnarray}}
\nc{\bra}[1]{\langle {#1}|}
\nc{\ket}[1]{|{#1}\rangle}
\nc{\ketw}[1]{({#1})^{\phantom{a}}_{{\cal W}}}
\nc{\chit}{\raisebox{0.25ex}{$\chi$}}
\nc{\chih}{\raisebox{0.25ex}{$\hat\chi$}}
\nc{\Db}{\mbox{\boldmath $D$}}
\nc{\Hb}{\mbox{\boldmath $H$}}
\nc{\calH}{{\cal H}}
\nc{\calR}{{\cal R}}
\nc{\calL}{{\cal L}}
\nc{\calV}{{\cal V}}
\nc{\Hc}{{\cal H}}
\nc{\Rc}{{\cal R}}
\nc{\Lc}{{\cal L}}
\nc{\Vc}{{\cal V}}
\nc{\Ib}{\mbox{\boldmath $I$}}
\nc{\qb}{\bar{q}}
\nc{\Ac}{\mathcal{A}}
\nc{\Bc}{\mathcal{B}}
\nc{\Cc}{\mathcal{C}}
\nc{\Dc}{\mathcal{D}}
\nc{\Ec}{\mathcal{E}}
\nc{\Gc}{\mathcal{G}}
\nc{\Ic}{\mathcal{I}}
\nc{\Jc}{\mathcal{J}}
\nc{\Oc}{\mathcal{O}}
\nc{\Pc}{\mathcal{P}}
\nc{\Sc}{\mathcal{S}}
\nc{\Tc}{\mathcal{T}}
\nc{\Wc}{\mathcal{W}}
\nc{\Xc}{\mathcal{X}}
\nc{\Yc}{\mathcal{Y}}
\nc{\Zc}{\mathcal{Z}}
\nc{\fus}{\mbox{}\,\hat\otimes\,\mbox{}}
\nc{\Pch}{\hat{\Pc}}
\nc{\Rch}{\hat{\Rc}}
\nc{\Dh}{\hat{\Delta}}
\nc{\rh}{\hat{r}}
\nc{\sh}{\hat{s}}
\nc{\taub}{\bar{\tau}}
\nc{\Jcb}{\Jc_{\mathrm{b}}}
\nc{\rtt}{\mathtt{r}}
\nc{\stt}{\mathtt{s}}
\nc{\cosR}{\cos\frac{\pi p'rr'}{p}}
\nc{\cosS}{\cos\frac{\pi pss'}{p'}}
\nc{\sinR}{\sin\frac{\pi p'rr'}{p}}
\nc{\sinS}{\sin\frac{\pi pss'}{p'}}
\def\vvdots{\mathinner{\mkern1mu\raise1pt\vbox{\kern7pt\hbox{.}}\mkern2mu
  \raise4pt\hbox{.}\mkern2mu\raise7pt\hbox{.}\mkern1mu}}
\nc{\gauss}[2]{\big[\!\!\begin{array}{c} {#1}\\ {#2} \end{array}\!\!\big]}
\nc{\bgauss}[2]{\Big[\!\!\begin{array}{c} {#1}\\ {#2} \end{array}\!\!\Big]}
\nc{\sbin}[2]{\left\{\!\!\!\begin{array}{c} {#1}\\ {#2}
\end{array}\!\!\!\right\}}
\nc{\sbinlr}[2]{\Big\langle\!\!\begin{array}{c} {#1}\\ {#2}
\end{array}\!\!\Big\rangle}
\nc{\bino}[2]{\left(\!\!\begin{array}{c} {#1}\\ {#2} \end{array}\!\!\right)}
\def\half {\mbox{$\textstyle \frac{1}{2}$}}

\definecolor{lightblue}{rgb}{.61,.61,1}
\definecolor{midblue}{rgb}{.7,.7,1}
\definecolor{lightlightblue}{rgb}{.85,.85,1}
\definecolor{lightestblue}{rgb}{.96,.96,1}
\definecolor{lightpurple}{rgb}{1,.65,1}
\nc{\ch}{{\rm ch}}
\nc{\R}{{\cal R}}
\nc{\dkk}{\delta_{j,\{k,k'\}}^{(2)}}
\nc{\drr}{\delta_{j,\{r,r'\}}^{(2)}}
\nc{\ddkk}{\delta_{j,\{k,k'\}}^{(4)}}
\nc{\dddkk}{\delta_{j,\{k,k'\}}^{(8)}}
\nc{\dnn}{\delta_{j,\{n,n'\}}^{(2)}}
\nc{\ddnn}{\delta_{j,\{n,n'\}}^{(4)}}
\nc{\dddnn}{\delta_{j,\{n,n'\}}^{(8)}}

\definecolor{pink}{rgb}{1,.65,.65}

\thicklines

\def\face#1#2#3#4{\ \
\begin{pspicture}[shift=-.6](0,-.25)(1,1.25)
\pspolygon[linewidth=.5pt,fillstyle=solid,fillcolor=lightlightblue](0,0)(1,0)(1,1)(0,1)(0,0)
\rput[tr](0,0){\scriptsize $#1$}
\rput[tl](1,0){\scriptsize $#2$}
\rput[bl](1,1){\scriptsize $#3$}
\rput[br](0,1){\scriptsize $#4$}
\end{pspicture}}

\def\sqfaceI#1#2#3#4{\ \
\begin{pspicture}[shift=-.6](0,-.25)(1,1.25)
\pspolygon[linewidth=.5pt,fillstyle=solid,fillcolor=lightlightblue](0,0)(1,0)(1,1)(0,1)(0,0)
\rput[tr](0,0){\scriptsize $#1$}
\rput[tl](1,0){\scriptsize $#2$}
\rput[bl](1,1){\scriptsize $#3$}
\rput[br](0,1){\scriptsize $#4$}
\psarc[linewidth=.5pt](0,1){.5}{270}{0}
\psarc[linewidth=.5pt](1,0){.5}{90}{180}
\psline[linewidth=1pt,linecolor=blue](0,0)(1,1)
\end{pspicture}}

\def\sqfaceE#1#2#3#4{\ \
\begin{pspicture}[shift=-.6](0,-.25)(1,1.25)
\pspolygon[linewidth=.5pt,fillstyle=solid,fillcolor=lightlightblue](0,0)(1,0)(1,1)(0,1)(0,0)
\rput[tr](0,0){\scriptsize $#1$}
\rput[tl](1,0){\scriptsize $#2$}
\rput[bl](1,1){\scriptsize $#3$}
\rput[br](0,1){\scriptsize $#4$}
\psarc[linewidth=.5pt](0,0){.5}{0}{90}
\psarc[linewidth=.5pt](1,1){.5}{180}{270}
\psline[linewidth=1pt,linecolor=blue](0,1)(1,0)
\end{pspicture}}

\nc{\Wtt}[6]{#1\!\!\left.\left(
\begin{matrix}
#5 & #4\\
#2 & #3
\end{matrix}
\right|#6\right)}
\nc{\thf}{s}
\nc{\lam}{\lambda}
\nc{\round}[1]{\left(#1\right)}
\nc{\therm}{t}
\def\floor#1{\lfloor #1\rfloor}
\def\bfloor#1{\big\lfloor #1\big\rfloor}

\begin{document}

\topmargin -5mm \oddsidemargin 5mm

\setcounter{page}{1}

\vspace{0mm}
\begin{center}
{\huge {\bf Off-Critical Logarithmic Minimal Models}}

\vspace{7mm}
{\Large Paul A. Pearce}\\[.3cm]
{\em Department of Mathematics and Statistics, University of Melbourne}\\
{\em Parkville, Victoria 3010, Australia}\\[.4cm]
\vspace{2mm}
{\Large Katherine A. Seaton}\\[.3cm]
{\em Department of Mathematics and Statistics, La Trobe University}\\
{\em Victoria 3086, Australia}\\[.4cm]
\end{center}

\vspace{4mm}
\centerline{{\bf{Abstract}}}
\vskip.3cm
\noindent
We consider the integrable minimal models ${\cal M}(m,m';t)$, corresponding to the $\varphi_{1,3}$ perturbation off-criticality, in the {\it logarithmic limit\,} $m, m'\to\infty$, $m/m'\to p/p'$ where $p, p'$ are coprime and the limit is taken through coprime values of $m,m'$. We view these off-critical minimal models ${\cal M}(m,m';t)$ as the continuum scaling limit of the Forrester-Baxter Restricted Solid-On-Solid (RSOS) models on the square lattice. Applying Corner Transfer Matrices to the Forrester-Baxter RSOS models in Regime~III, we argue that taking first the thermodynamic limit and second the {\it logarithmic limit\,} yields off-critical logarithmic minimal models ${\cal LM}(p,p';t)$ corresponding to the $\varphi_{1,3}$ perturbation of the critical logarithmic minimal models  ${\cal LM}(p,p')$. Specifically, in accord with the Kyoto correspondence principle, we show that the logarithmic limit of the one-dimensional configurational sums yields finitized quasi-rational characters of the Kac representations of the critical logarithmic minimal models  ${\cal LM}(p,p')$. We also calculate the  logarithmic limit  of certain off-critical observables ${\cal O}_{r,s}$ related to One Point Functions and show that the associated critical exponents $\beta_{r,s}=(2-\alpha)\,\Delta_{r,s}^{p,p'}$ produce all conformal dimensions $\Delta_{r,s}^{p,p'}<{(p'-p)(9p-p')\over 4pp'}$ in the infinitely extended Kac table. The corresponding Kac labels $(r,s)$ satisfy  $(p s-p' r)^2< 8p(p'-p)$. The exponent $2-\alpha ={p'\over 2(p'-p)}$ is obtained from the logarithmic limit of the free energy giving the conformal dimension $\Delta_t={1-\alpha\over 2-\alpha}={2p-p'\over p'}=\Delta_{1,3}^{p,p'}$ for the perturbing field $t$. As befits a non-unitary theory, some observables ${\cal O}_{r,s}$ diverge at criticality.

\renewcommand{\thefootnote}{\arabic{footnote}}
\setcounter{footnote}{0}

\tableofcontents

\newpage
\section{Introduction}

Consider the minimal models ${\cal M}(m,m')$~\cite{BPZ84} with coprime integers satisfying $2\le m<m'$. These are rational Conformal Field Theories (CFTs) with central charges
\bea
 c=1-{6(m-m')^2\over mm'},\qquad 2\le m<m',\quad\mbox{$m,m'$ coprime}
\eea
The conformal weights and their associated Virasoro characters are
\begin{align}
\Delta_{r,s}^{m,m'}&={(rm'-sm)^2-(m-m')^2\over 4mm'},\quad 1\le r\le m-1,\  1\le s\le m'-1\\
\mbox{ch}^{m,m'}_{r,s}(q)&={q^{-c/24+\Delta_{r,s}^{m,m'}}\over
(q)_\infty}\!\!\! \sum_{k=-\infty}^\infty \big[q^{k(k m
m'+rm'-sm)}-q^{(km+r)(km'+s)}\big] \label{vira}
\end{align}
where
\bea
(q)_n =
\prod_{k=1}^{n}(1-q^k),\qquad (q)_0=1,\qquad (q)_\infty =
\prod_{k=1}^{\infty}(1-q^k) \label{qn}
\eea
In these expressions,
$q=e^{\pi i\tau}$ is the modular nome.

The ``{\it logarithmic limit\/}"~\cite{Rasmussen} of the minimal CFTs is given symbolically by
\bea
\lim_{m,m'\to\infty, \  {m\over m'}\to {p\over p'}} {\cal M}(m,m')={\cal LM}(p,p'),\qquad 1\le p<p',\quad \mbox{$p,p'$ coprime}\label{logLimit}
\eea
where the limit is taken through coprime pairs $(m,m')$. Since $p\ge 1$, the limit must ultimately be taken through a sequence of non-unitary models with $m\ne m'-1$.
Here ${\cal LM}(p,p')$ denotes the logarithmic minimal models~\cite{PRZ06}.
The limit is taken directly in the continuum scaling limit, after the thermodynamic limit.
The equality indicates the identification of the spectra of these CFTs. In principle, the Jordan cells appearing in the reducible yet indecomposable representations of the logarithmic minimal models should emerge in this limit but there are subtleties~\cite{Rasmussen}.
Here we only consider the limit of spectra for which purpose the logarithmic limit is robust in the sense that it is independent of the choice of the sequence.

Taking the logarithmic limit of the conformal data of the rational minimal models ${\cal M}(m,m')$ yields directly~\cite{Rasmussen} the conformal data of the logarithmic minimal models ${\cal LM}(p,p')$
\bea
\begin{array}{rcll}
\disp c&\!\!=\!\!&\disp 1-{6(p-p')^2\over pp'}, &1\le p<p',\quad\ \ \mbox{$p,p'$ coprime}\\[12pt]
\disp \Delta_{r,s}^{p,p'}&\!\!=\!\!&\disp{(rp'-sp)^2-(p-p')^2\over 4pp'},\qquad &r=1,2,\ldots;\  s=1,2,\ldots\\[12pt]
\chi^{p,p'}_{r,s}(q)&\!\!=\!\!&\disp q^{-c/24+\Delta_{r,s}^{p,p'}}
{(1-q^{rs})\over (q)_\infty},\qquad  &r=1,2,\ldots;\  s=1,2,\ldots
\end{array}
\eea
The logarithmic characters are the quasi-rational characters of the so-called Kac representations which are organized into infinitely extended Kac tables. The Kac tables of conformal weights for critical dense polymers ${\cal LM}(1,2)$ and critical percolation ${\cal LM}(2,3)$ are shown in Table~1.
\def\myframe{\psframe[linewidth=0pt,fillstyle=solid,fillcolor=lightpurple]}
\begin{table}[htbp]
\begin{center}
\psset{unit=.7cm}
{\vspace{0in}
{\small
\begin{pspicture}(0,0)(7,11)
\psframe[linewidth=0pt,fillstyle=solid,fillcolor=lightlightblue](0,0)(7,11)
\multiput(0,0)(0,2){5}{\psframe[linewidth=0pt,fillstyle=solid,fillcolor=midblue](0,1)(7,2)}
\myframe(0,0)(1,4)
\myframe(1,1)(2,6)
\myframe(2,3)(3,8)
\myframe(3,5)(4,10)
\myframe(4,7)(5,11)
\myframe(5,9)(6,11)
\multirput(1,1)(1,0){6}{\pswedge[fillstyle=solid,fillcolor=red,linecolor=red](0,0){.25}{180}{270}}
\multirput(1,2)(1,0){6}{\pswedge[fillstyle=solid,fillcolor=red,linecolor=red](0,0){.25}{180}{270}}
\multirput(1,2)(0,2){5}{\pswedge[fillstyle=solid,fillcolor=red,linecolor=red](0,0){.25}{180}{270}}
\psgrid[gridlabels=0pt,subgriddiv=1]
\rput(.5,10.65){$\vdots$}\rput(1.5,10.65){$\vdots$}\rput(2.5,10.65){$\vdots$}\rput(3.5,10.65){$\vdots$}\rput(4.5,10.65){$\vdots$}\rput(5.5,10.65){$\vdots$}\rput(6.5,10.5){$\vvdots$}
\rput(.5,9.5){$\frac{63}8$}\rput(1.5,9.5){$\frac{35}8$}\rput(2.5,9.5){$\frac{15}8$}\rput(3.5,9.5){$\frac{3}8$}\rput(4.5,9.5){$-\frac 18$}\rput(5.5,9.5){$\frac{3}8$}\rput(6.5,9.5){$\cdots$}
\rput(.5,8.5){$6$}\rput(1.5,8.5){$3$}\rput(2.5,8.5){$1$}\rput(3.5,8.5){$0$}\rput(4.5,8.5){$0$}\rput(5.5,8.5){$1$}\rput(6.5,8.5){$\cdots$}
\rput(.5,7.5){$\frac{35}8$}\rput(1.5,7.5){$\frac {15}8$}\rput(2.5,7.5){$\frac 38$}\rput(3.5,7.5){$-\frac{1}8$}\rput(4.5,7.5){$\frac 38$}\rput(5.5,7.5){$\frac{15}8$}\rput(6.5,7.5){$\cdots$}
\rput(.5,6.5){$3$}\rput(1.5,6.5){$1$}\rput(2.5,6.5){$0$}\rput(3.5,6.5){$0$}\rput(4.5,6.5){$1$}\rput(5.5,6.5){$3$}\rput(6.5,6.5){$\cdots$}
\rput(.5,5.5){$\frac{15}8$}\rput(1.5,5.5){$\frac {3}{8}$}\rput(2.5,5.5){$-\frac 18$}\rput(3.5,5.5){$\frac{3}{8}$}\rput(4.5,5.5){$\frac {15}8$}\rput(5.5,5.5){$\frac{35}{8}$}\rput(6.5,5.5){$\cdots$}
\rput(.5,4.5){$1$}\rput(1.5,4.5){$0$}\rput(2.5,4.5){$0$}\rput(3.5,4.5){$1$}\rput(4.5,4.5){$3$}\rput(5.5,4.5){$6$}\rput(6.5,4.5){$\cdots$}
\rput(.5,3.5){$\frac 38$}\rput(1.5,3.5){$-\frac 18$}\rput(2.5,3.5){$\frac 38$}\rput(3.5,3.5){$\frac{15}8$}\rput(4.5,3.5){$\frac{35}8$}\rput(5.5,3.5){$\frac{63}8$}\rput(6.5,3.5){$\cdots$}
\rput(.5,2.5){$0$}\rput(1.5,2.5){$0$}\rput(2.5,2.5){$1$}\rput(3.5,2.5){$3$}\rput(4.5,2.5){$6$}\rput(5.5,2.5){$10$}\rput(6.5,2.5){$\cdots$}
\rput(.5,1.5){$-\frac 18$}\rput(1.5,1.5){$\frac 38$}\rput(2.5,1.5){$\frac{15}8$}\rput(3.5,1.5){$\frac{35}8$}\rput(4.5,1.5){$\frac{63}8$}\rput(5.5,1.5){$\frac{99}8$}\rput(6.5,1.5){$\cdots$}
\rput(.5,.5){$0$}\rput(1.5,.5){$1$}\rput(2.5,.5){$3$}\rput(3.5,.5){$6$}\rput(4.5,.5){$10$}\rput(5.5,.5){$15$}\rput(6.5,.5){$\cdots$}
{\color{blue}
\rput(.5,-.5){$1$}
\rput(1.5,-.5){$2$}
\rput(2.5,-.5){$3$}
\rput(3.5,-.5){$4$}
\rput(4.5,-.5){$5$}
\rput(5.5,-.5){$6$}
\rput(6.5,-.5){$r$}
\rput(-.5,.5){$1$}
\rput(-.5,1.5){$2$}
\rput(-.5,2.5){$3$}
\rput(-.5,3.5){$4$}
\rput(-.5,4.5){$5$}
\rput(-.5,5.5){$6$}
\rput(-.5,6.5){$7$}
\rput(-.5,7.5){$8$}
\rput(-.5,8.5){$9$}
\rput(-.5,9.5){$10$}
\rput(-.5,10.5){$s$}}
\end{pspicture}}}
\qquad
{\small
\qquad\qquad
\begin{pspicture}(0,0)(7,11)
\psframe[linewidth=0pt,fillstyle=solid,fillcolor=lightestblue](0,0)(7,11)
\psframe[linewidth=0pt,fillstyle=solid,fillcolor=lightblue](0,0)(1,2)
\psframe[linewidth=0pt,fillstyle=solid,fillcolor=lightlightblue](1,0)(2,11)
\psframe[linewidth=0pt,fillstyle=solid,fillcolor=lightlightblue](3,0)(4,11)
\psframe[linewidth=0pt,fillstyle=solid,fillcolor=lightlightblue](5,0)(6,11)
\psframe[linewidth=0pt,fillstyle=solid,fillcolor=lightlightblue](0,2)(7,3)
\psframe[linewidth=0pt,fillstyle=solid,fillcolor=lightlightblue](0,5)(7,6)
\psframe[linewidth=0pt,fillstyle=solid,fillcolor=lightlightblue](0,8)(7,9)
\multiput(0,0)(0,3){3}{\multiput(0,0)(2,0){3}{\psframe[linewidth=0pt,fillstyle=solid,fillcolor=midblue](1,2)(2,3)}}
\myframe(0,0)(1,3)
\myframe(1,1)(2,4)
\myframe(2,2)(3,6)
\myframe(3,4)(4,7)
\myframe(4,5)(5,9)
\myframe(5,7)(6,10)
\myframe(6,8)(7,11)
\multirput(2,1)(2,0){3}{\pswedge[fillstyle=solid,fillcolor=red,linecolor=red](0,0){.25}{180}{270}}
\multirput(2,2)(2,0){3}{\pswedge[fillstyle=solid,fillcolor=red,linecolor=red](0,0){.25}{180}{270}}
\multirput(2,3)(2,0){3}{\pswedge[fillstyle=solid,fillcolor=red,linecolor=red](0,0){.25}{180}{270}}
\multirput(1,3)(0,3){3}{\pswedge[fillstyle=solid,fillcolor=red,linecolor=red](0,0){.25}{180}{270}}
\multirput(2,3)(0,3){3}{\pswedge[fillstyle=solid,fillcolor=red,linecolor=red](0,0){.25}{180}{270}}
\psgrid[gridlabels=0pt,subgriddiv=1]
\rput(.5,10.65){$\vdots$}\rput(1.5,10.65){$\vdots$}\rput(2.5,10.65){$\vdots$}\rput(3.5,10.65){$\vdots$}\rput(4.5,10.65){$\vdots$}\rput(5.5,10.65){$\vdots$}\rput(6.5,10.5){$\vvdots$}
\rput(.5,9.5){$12$}\rput(1.5,9.5){$\frac{65}8$}\rput(2.5,9.5){$5$}\rput(3.5,9.5){$\frac{21}8$}\rput(4.5,9.5){$1$}\rput(5.5,9.5){$\frac{1}8$}\rput(6.5,9.5){$\cdots$}
\rput(.5,8.5){$\frac{28}3$}\rput(1.5,8.5){$\frac{143}{24}$}\rput(2.5,8.5){$\frac{10}3$}\rput(3.5,8.5){$\frac{35}{24}$}\rput(4.5,8.5){$\frac 13$}\rput(5.5,8.5){$-\frac{1}{24}$}\rput(6.5,8.5){$\cdots$}
\rput(.5,7.5){$7$}\rput(1.5,7.5){$\frac {33}8$}\rput(2.5,7.5){$2$}\rput(3.5,7.5){$\frac{5}8$}\rput(4.5,7.5){$0$}\rput(5.5,7.5){$\frac{1}8$}\rput(6.5,7.5){$\cdots$}
\rput(.5,6.5){$5$}\rput(1.5,6.5){$\frac {21}8$}\rput(2.5,6.5){$1$}\rput(3.5,6.5){$\frac{1}8$}\rput(4.5,6.5){$0$}\rput(5.5,6.5){$\frac{5}8$}\rput(6.5,6.5){$\cdots$}
\rput(.5,5.5){$\frac{10}3$}\rput(1.5,5.5){$\frac {35}{24}$}\rput(2.5,5.5){$\frac 13$}\rput(3.5,5.5){$-\frac{1}{24}$}\rput(4.5,5.5){$\frac 13$}\rput(5.5,5.5){$\frac{35}{24}$}\rput(6.5,5.5){$\cdots$}
\rput(.5,4.5){$2$}\rput(1.5,4.5){$\frac 58$}\rput(2.5,4.5){$0$}\rput(3.5,4.5){$\frac{1}8$}\rput(4.5,4.5){$1$}\rput(5.5,4.5){$\frac{21}8$}\rput(6.5,4.5){$\cdots$}
\rput(.5,3.5){$1$}\rput(1.5,3.5){$\frac 18$}\rput(2.5,3.5){$0$}\rput(3.5,3.5){$\frac{5}8$}\rput(4.5,3.5){$2$}\rput(5.5,3.5){$\frac{33}8$}\rput(6.5,3.5){$\cdots$}
\rput(.5,2.5){$\frac 13$}\rput(1.5,2.5){$-\frac 1{24}$}\rput(2.5,2.5){$\frac 13$}\rput(3.5,2.5){$\frac{35}{24}$}\rput(4.5,2.5){$\frac{10}3$}\rput(5.5,2.5){$\frac{143}{24}$}\rput(6.5,2.5){$\cdots$}
\rput(.5,1.5){$0$}\rput(1.5,1.5){$\frac 18$}\rput(2.5,1.5){$1$}\rput(3.5,1.5){$\frac{21}8$}\rput(4.5,1.5){$5$}\rput(5.5,1.5){$\frac{65}8$}\rput(6.5,1.5){$\cdots$}
\rput(.5,.5){$0$}\rput(1.5,.5){$\frac 58$}\rput(2.5,.5){$2$}\rput(3.5,.5){$\frac{33}8$}\rput(4.5,.5){$7$}\rput(5.5,.5){$\frac{85}8$}\rput(6.5,.5){$\cdots$}
{\color{blue}
\rput(.5,-.5){$1$}
\rput(1.5,-.5){$2$}
\rput(2.5,-.5){$3$}
\rput(3.5,-.5){$4$}
\rput(4.5,-.5){$5$}
\rput(5.5,-.5){$6$}
\rput(6.5,-.5){$r$}
\rput(-.5,.5){$1$}
\rput(-.5,1.5){$2$}
\rput(-.5,2.5){$3$}
\rput(-.5,3.5){$4$}
\rput(-.5,4.5){$5$}
\rput(-.5,5.5){$6$}
\rput(-.5,6.5){$7$}
\rput(-.5,7.5){$8$}
\rput(-.5,8.5){$9$}
\rput(-.5,9.5){$10$}
\rput(-.5,10.5){$s$}}
\end{pspicture}}
\end{center}
\caption{Infinitely extended Kac tables of conformal weights $\Delta_{r,s}^{p,p'}$ for critical dense polymers ${\cal LM}(1,2)$ and critical percolation ${\cal LM}(2,3)$. The positions corresponding to irreducible representations are marked with a quadrant in the top-right corner.
The boxes of the conformal dimensions \mbox{$\Delta_{r,s}< {(p'-p)(9p-p')\over 4p p'}$} that are accessible, using Corner Transfer Matrices to calculate off-critical observables, are shaded in pink and occur in the band where $(r,s)$ satisfy $(p s-p' r)^2< 8p(p'-p)$.}
\end{table}

\begin{figure}[b]
\psset{unit=1.05cm}
\begin{center}
\begin{pspicture}(0,0)(4,2)
\rput(0,0){${\cal LM}(p,p')$}
\rput(4,0){$\;{\cal LM}(p,p';t)$}
\rput(0,1.5){${\cal M}(m,m')$}
\rput(4,1.5){$\;{\cal M}(m,m';t)$}
\rput(1.8,.3){$t\sim \varphi_{1,3}$}
\rput(1.8,1.8){$t\sim \varphi_{1,3}$}
\rput(-.4,.9){\mbox{log}}
\rput(3.6,.9){\mbox{log}}
\psline[linewidth=.5pt,arrowsize=7pt]{->}(1,0)(3,0)
\psline[linewidth=.5pt,arrowsize=7pt]{->}(1,1.5)(3,1.5)
\psline[linewidth=.5pt,arrowsize=7pt]{->}(0,1.2)(0,.3)
\psline[linewidth=.5pt,arrowsize=7pt]{->}(4,1.2)(4,.3)
\end{pspicture}
\end{center}
\caption{Commutative diagram of the logarithmic limit and $\varphi_{1,3}$ off-critical perturbation.}
\end{figure}

In this paper, we consider the {\it logarithmic limit\/} of the
off-critical minimal models and argue that it provides an integrable
off-critical perturbation of the logarithmic minimal models. We denote the off-critical models by $ {\cal M}(m,m';t)$ or $ {\cal M}(m,m';q)$ where
$t=e^{-\epsilon}$ and $q=e^{-4\pi^2 (m'-m)/m' \epsilon}$ are conjugate off-critical nomes measuring the temperature.
Symbolically, as in Figure~1, we write
\bea
\lim_{m,m'\to\infty, \  {m\over m'}\to {p\over p'}} {\cal M}(m,m';\therm)={\cal LM}(p,p';\therm),\qquad
1\le p<p',\quad \mbox{$p,p'$ coprime} \label{offlogLimit}
\eea
where now $\therm$ is an
elliptic nome measuring the departure from criticality.
Since degeneracies necessary for the formation of Jordan cells are lifted
off-criticality, the off-critical logarithmic minimal models are not expected to
exhibit the  {\it logarithmic} structures, such as reducible yet
indecomposable representations, associated with logarithmic CFTs.
Nevertheless, we call these models {\it off-critical logarithmic
minimal models} to emphasize their relation to their logarithmic
critical counterparts. Notice that although ${\cal LM}(2,3)={\cal
LM}(2,3;0)$ is critical bond percolation, the integrable
off-critical percolation model ${\cal LM}(2,3;\therm)$ does not
coincide with the off-critical bond percolation model. To put it another way, as shown in Appendix~A, the off-critical model
${\cal LM}(2,3;\therm)$ is the integrable $\varphi_{1,3}$ perturbation
of critical bond percolation
\bea
\Delta_{t}=\frac{1-\alpha}{2-\alpha}=\frac{1}{3}=\Delta_{1,3}^{2,3},\qquad t\sim\varphi_{1,3}
\eea
In contrast, the usual off-critical bond percolation model~\cite{Stauffer}
corresponds to a $\varphi_{2,1}$ perturbation. The (mean cluster number) critical exponent is $\alpha_p=-\frac{2}{3}$
corresponding to the off-critical $\varphi_{2,1}$ perturbation
\bea
\Delta_{p}=\frac{1-\alpha_p}{2-\alpha_p}=\frac{5}{8}=\Delta_{2,1}^{2,3},\qquad p\sim\varphi_{2,1}
\eea
where here $p$ is the bond occupation probability with $p_c=\half$~\cite{Kesten}.

\goodbreak
The content of the paper is as follows. It is known that the
rational minimal models are described by the continuum scaling limit
of the Forrester-Baxter RSOS models~\cite{ABF84,FB84} in Regime~III.
In Section~2, we use Corner Transfer Matrix (CTM)
techniques to study the logarithmic limit of the one-dimensional
configurational sums of the Forrester-Baxter RSOS models in
Regime~III. We show that, up to the leading terms involving the
central charges and conformal dimensions, the one-dimensional
configurational sums reproduce the finitized quasi-rational Kac character
formulas. This observation is in agreement with the general {\it
correspondence principle} of the Kyoto school~\cite{Kyoto} which is
valid in Regime~III. In Section 3, we obtain general expressions for
the One Point Functions (OPFs), first for the Forrester-Baxter models and then in the logarithmic limit.
We extract critical exponents from the behaviour of suitably defined Generalized Order Parameters
near the critical point $t\to 0$.
In the logarithmic limit, these exponents are simply related to conformal dimensions in the
infinitely extended Kac table. In Appendix~A, we look at the behaviour of the
logarithmic limit of the free energy as $\therm \to 0$ and
argue that the relevant perturbation, away from the critical logarithmic minimal models,
is the thermal $\varphi_{1,3}$ perturbation. In Appendix~B, we collect relevant properties of elliptic functions.

\section{1-$d$ Configurational Sums of Limiting Forrester-Baxter Models}
\subsection{Forrester-Baxter models}

\setlength{\unitlength}{28pt}
\psset{unit=28pt}
The Forrester-Baxter RSOS lattice models~\cite{FB84} are defined on a square lattice with heights
{$a=1,2,\ldots,m'-1$} restricted so that nearest neighbour heights differ by $\pm 1$. The heights thus live on the $A_{m'-1}$ Dynkin diagram.
The Boltzmann weights are
\begin{align}
&\Wtt{W}{a}{a\mp1}{a}{a\pm1}u=\frac{\thf(\lam-u)}{\thf(\lam)}\ \ \face {a}{a\!\mp\!1}{a}{a\!\pm\!1}\ \;=\ \;
\frac{\thf(\lam-u)}{\thf(\lam)}\ \ \sqfaceI a{a\!\mp\!1}a{a\!\pm\!1}\label{Boltzmann1}\\
&\Wtt{W}{a\mp1}{a}{a\pm1}{a}u=
{g_{a\mp 1}\over g_{a\pm 1}}\,
\frac{\thf((a\pm 1)\lam)}{\thf(a\lam)}\,
\frac{\thf(u)}{\thf(\lam)}\ \ \ \face {a\!\mp\!1}{a}{a\!\pm\!1}{a}\ \; =\ \;
{g_{a\mp 1}\over g_{a\pm 1}}\,
\frac{\thf((a\pm 1)\lam)}{\thf(a\lam)}\,
\frac{\thf(u)}{\thf(\lam)}\ \ \;
\sqfaceE {a\!\mp\!1}a{a\!\pm\!1}a\label{Boltzmann2}\\
&\Wtt{W}{a\pm1}{a}{a\pm1}{a}u=
\frac{\thf(a\lam\pm u)}{\thf(a\lam)}\ \;\face {a\!\pm\!1}{a}{a\!\pm\!1}{a}\nonumber\\
&\quad=\;{c(0)c((a\!\pm\!1)\lam\!\pm\! u)\over c((a\!\pm\!1)\lam)c(u)}{\thf(\lam\!-\!u)\over\thf(\lam)}
\ \;\sqfaceI {a\!\pm\!1}{a}{a\!\pm\!1}{a}\ \;
+{c(\lam)c(a\lam\!\pm\! u)\over c((a\!\pm\!1)\lam)c(u)}{\thf((a\!\pm\!1)\lam)\over \thf(a\lam)}\,{\thf(u)\over\thf(\lam)}
\ \ \;\sqfaceE {a\!\pm\!1}{a}{a\!\pm\!1}{a}\label{Boltzmann3}
\end{align}
Here $\thf(u)=\vartheta_1(u,\therm)$ and $c(u)=\vartheta_4(u,\therm)$ are standard elliptic theta
functions~\cite{GR} as in Appendix~B,
$u$ is the spectral parameter and the elliptic nome $\therm=e^{-\epsilon}$
is a temperature-like variable, with $\therm^2$ measuring the
departure from criticality corresponding to the $\varphi_{1,3}$
integrable perturbation. The gauge factors $g_a$ are inessential and
can be set either to $1$ or $g_a=\sqrt{s(a\lambda)}$, as in \cite{FB84}, to restore
reflection symmetry. The crossing parameter $\lambda$ is
\begin{equation}
\lambda={(m'-m)\pi\over m'},\qquad 2\le m<m',\qquad\mbox{$m,m'$ coprime}
\label{crossing}
\end{equation}

The RSOS faces are decomposed into decorated faces or tiles (2-tangles) of the planar algebra~\cite{planarJones} having height degrees of freedom at the corners and twofold internal degrees of freedom associated with the direction of the loop segments, or equivalently, the diagonal bonds.
These models can be viewed as generalized models of polymers and percolation with percolation properties described in terms of percolating loops or percolating bond clusters. Sites in a common bond cluster share a common height $a$.
Integrability derives from the
fact that the local face weights satisfy the Yang-Baxter equation~\cite{BaxBook}. In the symmetric gauge, the face weights also obey the crossing symmetry
\begin{equation}
\Wtt{W}{a}{b}{c}{d}{\lambda-u}=\sqrt{\frac{s(a\lambda)s(c\lambda)}{s(b\lambda)s(d\lambda)}}\,\Wtt{W}{d}{a}{b}{c}{u}
\label{cross}
\end{equation}

At criticality and in the symmetric gauge, the Boltzmann weights reduce to
\bea
\Wtt{W}abcdu
=\;{\thf(\lam\!-\!u)\over\thf(\lam)}\,\delta_{a,c}\;
+\sqrt{{\thf(a\lam)\thf(c\lam)\over \thf(b\lam)\thf(d\lam)}}\,{\thf(u)\over\thf(\lam)}\,\delta_{b,d}\label{heightRep}
\eea
with $s(u)=\sin u$. The height dependent square root factors give rise to a matrix representation~\cite{OwczBaxt} of the Temperley-Lieb~\cite{TempLieb} generators $e_j$ acting on paths built on the graph $A_{m'-1}$. It follows that the associated face operators are
\bea
X_j(u)={\thf(\lam\!-\!u)\over\thf(\lam)}\,I+{\thf(u)\over\thf(\lam)}\,e_j\label{faceOp}
\eea
where $I$ is the identity matrix. The Temperley-Lieb generators satisfy $e_j^2=\beta e_j$ where the loop fugacity
$\beta=2\cos{(m'-m)\pi\over m'}\to 2\cos{(p'-p)\pi\over p'}$ in the logarithmic limit. The face operators (\ref{faceOp}) thus coincide with those of the logarithmic minimal models ${\cal LM}(p,p')$ except that in \cite{PRZ06} the loop representation is used as opposed to the height representation of the Temperley-Lieb algebra used in this paper. On the strip, it is possible to construct a Markov trace~\cite{JonesMarkov} on the Temperley-Lieb algebra such that the traces of words in the algebra are independent of the choice of representation. It follows that, if we use this Markov trace on the strip, the resulting partition functions for the logarithmic theories ${\cal LM}(p,p')$ will agree with the logarithmic limit of the partition functions of the RSOS theories (\ref{heightRep}).

It does not make sense to take the logarithmic limit (\ref{offlogLimit}) of the face weights directly because the factors $s(a\lambda)$ vanish whenever $a$ is a multiple of $p'$ causing some weights to vanish and others to diverge.
To avoid this problem, we take the logarithmic limit after the thermodynamic limit. In practice, however, quantities for
${\cal LM}(p,p';t)$ can be obtained within arbitrary precision by considering a non-unitary minimal model with Boltzmann weights (\ref{Boltzmann1})--(\ref{Boltzmann3}) and judiciously choosing the crossing parameter, say $\lam=2,000,000\,\pi/3,000,001$ for off-critical percolation.

\begin{figure}[htb]
\psset{unit=.7cm}
\begin{center}
\begin{pspicture}(0,-.5)(10,10.5)
\psframe[linewidth=0pt,fillstyle=solid,fillcolor=lightlightblue](4,0)(6,10)
\psframe[linewidth=0pt,fillstyle=solid,fillcolor=lightlightblue](3,1)(7,9)
\psframe[linewidth=0pt,fillstyle=solid,fillcolor=lightlightblue](2,2)(8,8)
\psframe[linewidth=0pt,fillstyle=solid,fillcolor=lightlightblue](1,3)(9,7)
\psframe[linewidth=0pt,fillstyle=solid,fillcolor=lightlightblue](0,4)(10,6)
\psline[linewidth=1.5pt](5,0)(5,10)
\psline[linewidth=1.5pt](0,5)(10,5)
\psframe[linewidth=.5pt](4,0)(6,10)
\psframe[linewidth=.5pt](3,1)(7,9)
\psframe[linewidth=.5pt](2,2)(8,8)
\psframe[linewidth=.5pt](1,3)(9,7)
\psframe[linewidth=.5pt](0,4)(10,6)
\rput[tl](8.25,1.75){\Large$A$}
\rput[bl](8.25,8.25){\Large$B$}
\rput[br](1.75,8.25){\Large$C$}
\rput[tr](1.75,1.75){\Large$D$}
\rput[bl](5.1,5.15){\Large$S$}
\rput[tr](4.9,4.9){\Large$\sigma_0$}
\multirput(0,0)(1,1){4}{\rput[tl](6,1){$\tilde{b}$}}
\multirput(0,0)(-1,1){4}{\rput[bl](9,6){$\tilde{b}$}}
\multirput(0,0)(-1,-1){4}{\rput[br](4,9){$\tilde{b}$}}
\multirput(0,0)(1,-1){4}{\rput[tr](1,4){$\tilde{b}$}}
\multirput(0,0)(1,1){5}{\rput[tl](6,0){$\tilde{c}$}}
\multirput(0,0)(-1,1){5}{\rput[bl](10,6){$\tilde{c}$}}
\multirput(0,0)(-1,-1){5}{\rput[br](4,10){$\tilde{c}$}}
\multirput(0,0)(1,-1){5}{\rput[tr](0,4){$\tilde{c}$}}
\rput[l](10,5){$\tilde{b}$}
\rput[b](5,10){$\tilde{b}$}
\rput[r](0,5){$\tilde{b}$}
\rput[t](5,0){$\tilde{b}$}
\rput(5,1){\pscircle*{2pt}}
\multiput(4,2)(1,0){3}{\rput(0,0){\pscircle*{2pt}}}
\multiput(3,3)(1,0){5}{\rput(0,0){\pscircle*{2pt}}}
\multiput(2,4)(1,0){7}{\rput(0,0){\pscircle*{2pt}}}
\multiput(1,5)(1,0){9}{\rput(0,0){\pscircle*{2pt}}}
\multiput(2,6)(1,0){7}{\rput(0,0){\pscircle*{2pt}}}
\multiput(3,7)(1,0){5}{\rput(0,0){\pscircle*{2pt}}}
\multiput(4,8)(1,0){3}{\rput(0,0){\pscircle*{2pt}}}
\multiput(5,9)(1,0){1}{\rput(0,0){\pscircle*{2pt}}}
\end{pspicture}
\caption{Square lattice geometry for Corner Transfer Matrices $A,B,C,D$. The half-row configurations are given by $\sigma=\{\sigma_0,\sigma_1,\sigma_2,\ldots,\sigma_N,\sigma_{N+1}\}$ with the boundary conditions $\sigma_N=\tilde{b}$ and $\sigma_{N+1}=\tilde{c}$ with $\tilde{b},\tilde{c}=1,2,\ldots,m'-1$ and $\tilde{b}-\tilde{c}=\pm 1$.
The center height $\sigma_0=1,2,\ldots,m'-1$ is fixed to the value $s$ by the diagonal matrix $S$ with entries $\langle \sigma|S|\sigma'\rangle=\delta(\sigma_0,s) \prod_{j=0}^{N+1} \delta(\sigma_j,\sigma_j')$. The condition $N+s-\tilde{b}=0$ mod 2 ensures that it is possible to get from $s$ to $\tilde{b}$ in $N$ steps. Internal heights, shown with a solid disk, are summed over. The groundstate boundary conditions are labelled by $r=1,2,\ldots, m-1$. If $N+s-b=0$ mod 2, the boundary heights are fixed as shown with $\tilde{b}=b=b(r)=\floor{{rm'\over m}}$ and $\tilde{c}=b+1$. Otherwise, if $N+s-b=1$ mod 2, the boundary heights are $\tilde{b}=b+1$ and $\tilde{c}=b=b(r)$. In both cases, $b=b(r)=\mbox{Min}[\tilde{b},\tilde{c}]$.}
\end{center}
\end{figure}
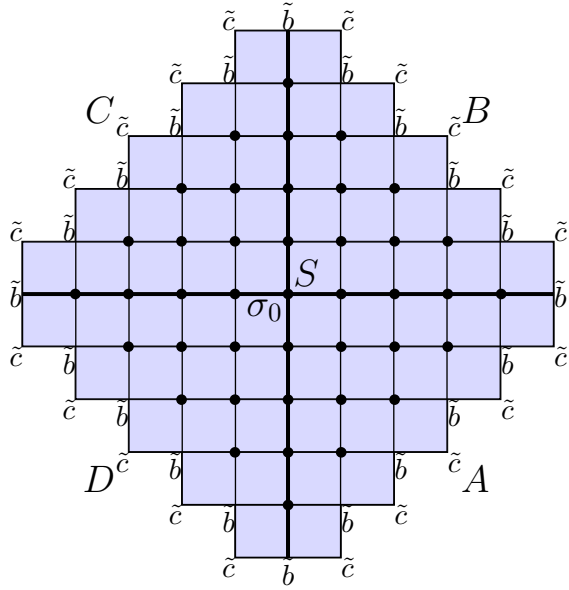

\subsection{Corner transfer matrices and 1-dimensional configurational sums}

The one-dimensional configurational sums~\cite{BaxBook} arise from the
diagonalization of the Corner Transfer Matrices (CTMs) of Figure~2. Allowing for
translations, the Forrester-Baxter models admit $2(m-1)$
groundstates in Regime~III wherein the heights outside of the finite region shown in Figure~2 alternate between $b$
and $b+1$ on the two sublattices of the square lattice. The allowed
values of $b$ are
\bea
b=b(r)=\bfloor{{r m'\over m}},\qquad
r=1,2,\ldots,m-1 \label{b(r)}
\eea
with inverse
\bea
r=r(b)=\bfloor{{b\,m\over m'}}+1 \label{r(b)}
\eea
In a given sector, specified by the boundary conditions $(\sigma_0,\sigma_N,\sigma_{N+1})$, the eigenvalues of the CTMs
are labelled by paths $\sigma$ on the $A_{m'-1}$ diagram that start
at $\sigma_0=a$ and end with either $\sigma_N=b$, $\sigma_{N+1}=b+1$
or $\sigma_N=b+1$, $\sigma_{N+1}=b$. We denote this set of paths by
${\cal P}_{a\tilde{b}\tilde{c}}^{m'}(N)$ or just ${\cal P}_{abc}^{m'}(N)$ when $b,c$ are generic dummy heights which should not cause confusion with the groundstate labels $b=b(r)$. A typical path with the groundstate bands
$(b,b+1)$ shaded is shown in Figure~3. More precisely, the
one-dimensional configurational sums are given by \bea
X_{abc}^{(N)}(q)\!=\!\disp\sum_\sigma q^{\sum_{j=1}^N j
H(\sigma_{j-1},\sigma_j,\sigma_{j+1})-E_0},\quad
\sigma\!=\!\{\sigma_0,\sigma_1,\ldots,\sigma_N,\sigma_{N+1}\}\in
{\cal P}_{abc}^{m'}(N) \label{oned}
\eea
subject to the boundary conditions
\bea
(\sigma_0,\sigma_N,\sigma_{N+1})\!=\!(a,b,c)\label{bc}
\eea
where the local energy function is
\bea
H(a,a\mp 1,a)=\pm
\bfloor{{a(m'-m)\over m'}},\qquad H(a-1,a,a+1)=H(a+1,a,a-1)=\half
\eea
As a consequence of the crossing symmetry (\ref{cross}), the conjugate nome
in the one-dimensional configurational sums (\ref{oned}) is
$q=e^{-4\pi\lambda/\epsilon}$ with the original nome of the face weights given by
$\therm=e^{-\epsilon}$.
The constant $E_0$ is the energy of the path
with minimum energy.  Its presence ensures that the $q$-series
begins as $1+O(q)$.
Explicitly, the ground state energy for boundary
conditions (\ref{bc}) is
\bea
E_0={1\over 4}(a-b)(a-c)+{1\over 2}\big[(a-c) +(c-b)(N+1)\big]\bfloor{\frac{c(m'-m)}{m'}}
\eea
The one-dimensional configurational sums are purely combinatorial objects~\cite{FodaW}. They are the energy generating functions for the CTM eigenvalues.
\begin{figure}[tbp]
\psset{unit=.6cm}
\begin{center}
\begin{pspicture}(0,.3)(10,6.5)
\psframe[linewidth=0pt,fillstyle=solid,fillcolor=lightlightblue](0,5)(10,6)
\psframe[linewidth=0pt,fillstyle=solid,fillcolor=lightestblue](0,4)(10,5)
\psframe[linewidth=0pt,fillstyle=solid,fillcolor=lightlightblue](0,3)(10,4)
\psframe[linewidth=0pt,fillstyle=solid,fillcolor=lightlightblue](0,2)(10,3)
\psframe[linewidth=0pt,fillstyle=solid,fillcolor=lightestblue](0,1)(10,2)
\psframe[linewidth=0pt,fillstyle=solid,fillcolor=lightlightblue](0,0)(10,1)
\multirput(1,0)(1,0){9}{\psline[linewidth=.5pt](0,0)(0,6)}
\psline[linewidth=1pt](0,0)(4,4)(5,3)(6,4)(7,3)(8,4)(9,3)(10,4)
\rput(0,-.4){\small $0$}
\rput(1,-.4){\small $1$}
\rput(2,-.4){\small $2$}
\rput(3,-.4){\small $3$}
\rput(4,-.4){\small $4$}
\rput(5,-.4){\small $5$}
\rput(6,-.4){\small $6$}
\rput(7,-.4){\small $7$}
\rput(8,-.4){\small $8$}
\rput(9,-.4){\small $9$}
\rput(10,-.4){\small $10$}
\rput(-.3,0){\small $1$}
\rput(-.3,1){\small $2$}
\rput(-.3,2){\small $3$}
\rput(-.3,3){\small $4$}
\rput(-.3,4){\small $5$}
\rput(-.3,5){\small $6$}
\rput(-.3,6){\small $7$}
\end{pspicture}
\end{center}
\caption{The path $\sigma=\{1,2,3,4,5,4,5,4,5,4,5\}$ of ${\cal M}(5,8;q)$ on $A_7$ with length $N=9$ and boundary condition
$(\sigma_0,\sigma_N,\sigma_{N+1})=(1,4,5)=(s,b,b+1)$. The groundstate bands $(b,b+1)$ at $b=b(r)=\bfloor{{r m'\over m}}=1,3,4,6$ are shown shaded. This is the path with minimum energy $E_0=6$ in the $(r,s)=(3,1)$ sector since $b(3)=4$.}
\end{figure}
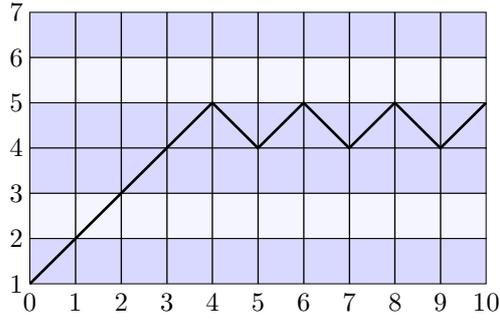

The one-dimensional configurational sums can be
evaluated~\cite{FB84} by solving recursion equations subject to
appropriate initial conditions.  Explicitly, for the Forrester-Baxter models, the
solutions are
\bea
\begin{array}{rcl}
X_{ab\{b+1\}}^{(N)}(q)&\!\!\!\!=\!\!\!\!&\disp\sum_{k=-\infty}^\infty \!\!
\Big(q^{k(k m m'+rm'-am)}\bgauss{N}{{N+a-b\over 2}-km'}-q^{(km+r)(km'+a)}\bgauss{N}{{N-a-b\over 2}-k m'}\Big)\\[18pt]
X_{a\{b+1\}b}^{(N)}(q)&\!\!\!\!=\!\!\!\!&\disp\sum_{k=-\infty}^\infty \!\!\!
\Big(q^{k(k m m'+rm'-am)}\bgauss{N}{{N+a-b-1\over 2}-km'}-q^{(km+r)(km'+a)}\bgauss{N}{{N-a-b-1\over 2}-k m'}\Big)
\end{array}\label{sums}
\eea
for the two mod $2$ parities of $N+a-b$ where $b=1,2,\ldots,m-2$.
The $q$-binomials appearing in these formulas are defined by
\bea
\bgauss{m+n}{m}
=\begin{cases}
\dfrac{(q)_{m+n}}{(q)_{m}(q)_{n}}, &\quad m,n\ge 0 \\[4pt]
0, &\quad \text{otherwise}\\
\end{cases}
\eea
The one-dimensional configurational sums coincide with finitized Virasoro characters
\bea
\mbox{ch}^{(N)}_{r,s}(q)=\begin{cases}
q^{-c/24+\Delta_{r,s}^{m,m'}} X_{sb\{b+1\}}^{(N)}(q),& b=b(r),\quad N+s-b=0\ \mbox{mod 2}\\[6pt]
q^{-c/24+\Delta_{r,s}^{m,m'}} X_{s\{b+1\}b}^{(N)}(q),& b=b(r),\quad N+s-b=1\ \mbox{mod 2}
\end{cases} \label{finvir}
\eea
where $q$ is now the modular nome and
\bea
\lim_{N\to\infty}
\mbox{ch}^{(N)}_{r,s}(q)=\mbox{ch}^{m,m'}_{r,s}(q),\qquad
r=1,2,\ldots,m-1;\quad  s=1,2,\ldots,m'-1
\eea

\subsection{Logarithmic limit of Forrester-Baxter models}
In the logarithmic limit, the heights $\sigma_j$ are unrestricted and live on the (one-sided) $A_\infty$ diagram. Notice however that, for given boundary conditions and any fixed $N$, the heights can only assume a finite range of values. In this limit, the banding pattern is repeated periodically with period $p'$. Explicitly, the groundstate bands $(b,b+1)$ are given by
\bea
b=b(r)=\bfloor{{rp'\over p}},\qquad r=1,2,3,\ldots
\eea
For the ${\cal LM}(1,p';q)$ theories, the groundstate bands are simply $b(r)=p' r$. For off-critical logarithmic dense polymers ${\cal LM}(1,2;q)$, percolation
${\cal LM}(2,3;q)$ and Lee-Yang ${\cal LM}(2,5;q)$ the groundstate bands are located at
\bea
b(r)=\begin{cases}
2,4,6,8,10,12,\ldots&(p,p')=(1,2)\\
1,3,4,6,7,9,10,12\ldots&(p,p')=(2,3)\\
2,5,7,10,12,15,17,20,\ldots&(p,p')=(2,5)
\end{cases}\qquad r=1,2,3,\ldots
\eea The one-dimensional configurational sums give a finite set of
eigenvalues of the infinite system. The $N$ in these formulas is a
{\it truncation size} and not the system size which has already been
taken to infinity. We can therefore apply the logarithmic limit
directly to the one-dimensional configurational sums.

The unrestricted one-dimensional configurational sums are given by
\bea
Y_{abc}^{(N)}(q)\!=\!\sum_\sigma q^{\sum_{j=1}^N j H^\infty(\sigma_{j-1},\sigma_j,\sigma_{j+1})-E_0^\infty},\quad
\sigma\!=\!\{\sigma_0,\sigma_1,\ldots,\sigma_N,\sigma_{N+1}\}\in {\cal P}_{abc}^{\infty}(N)
\eea
subject to the boundary conditions (\ref{bc}) with local energy function
\bea
H^\infty(a,a\mp 1,a)=\pm \bfloor{{a(p'-p)\over p'}},\qquad H^\infty(a-1,a,a+1)=H^\infty(a+1,a,a-1)=\half
\eea
Taking the logarithmic limit of the Forrester-Baxter results (\ref{sums}), gives
\bea
\begin{array}{rcl}
Y_{ab\{b+1\}}^{(N)}(q)&\!\!=\!\!&\disp
\bgauss{N}{{N+a-b\over 2}}-q^{ra}\bgauss{N}{{N-a-b\over 2}},\qquad\qquad N+a-b=0\ \mbox{mod 2}\\[10pt]
Y_{a\{b+1\}b}^{(N)}(q)&\!\!=\!\!&\disp
\bgauss{N}{{N+a-b-1\over 2}}-q^{ra}\bgauss{N}{{N-a-b-1\over 2}},\qquad N+a-b=1\ \mbox{mod 2}
\end{array}
\eea
In accord with the  {\it Kyoto correspondence principle}~\cite{Kyoto}, these one-dimensional configurational sums coincide with finitized quasi-rational characters of the Kac representations of the critical ${\cal LM}(p,p')$ models
\bea
\chi^{(N)}_{r,s}(q)=\begin{cases}
q^{-c/24+\Delta_{r,s}^{p,p'}} Y_{sb\{b+1\}}^{(N)}(q),& b=b(r),\quad N+s-b=0\ \mbox{mod 2}\\[6pt]
q^{-c/24+\Delta_{r,s}^{p,p'}} Y_{s\{b+1\}b}^{(N)}(q),& b=b(r),\quad N+s-b=1\ \mbox{mod 2}
\end{cases}\label{logfinchar}
\eea with  \bea \lim_{N\to
\infty}\chi^{(N)}_{r,s}(q)=\chi^{p,p'}_{r,s}(q),\qquad
r,s=1,2,3,\ldots
\eea
Again $q$ is the modular nome in these character formulas. An alternative view is that, in Regime~III, the massive-conformal renormalization group flow is uninteresting. In the words of Melzer~\cite{Melzer}, there exists a simple massive-conformal dictionary between eigenstates. This is not the case in Regime~IV.

We note that the one-dimensional configurational sums of the logarithmic theory
\bea
\bar{Y}_{abc}^{(N)}(q)=q^{E_0^\infty}Y_{abc}^{(N)}(q)
\eea
satisfy~\cite{Trott} CTM recursions of the usual form
\bea
\begin{array}{rcl}
\bar{Y}_{ab\{b+1\}}^{(N)}(q)\!\!&=&\!\!q^{N\floor{{(b+1)(p'-p)\over p'}}}\bar{Y}_{a\{b+1\}b}^{(N-1)}(q)+q^{N\over 2}\bar{Y}_{a\{b-1\}b}^{(N-1)}(q)\\[6pt]
\bar{Y}_{ab\{b-1\}}^{(N)}(q)\!\!&=&\!\!q^{N\over 2}\bar{Y}_{a\{b+1\}b}^{(N-1)}(q)+q^{-N\floor{{(b-1)(p'-p)\over p'}}}\bar{Y}_{a\{b-1\}b}^{(N-1)}(q)
\end{array}\label{logRecursion}
\eea
subject to the initial and boundary conditions
\bea
\bar{Y}_{abc}^{(0)}(q)=\delta_{a,b},\qquad \bar{Y}_{a\{0\}1}^{(N)}(q)=0
\eea
Here the groundstate energy $E_0^\infty$ for walks on $A^\infty$ is the logarithmic limit of the expression for $E_0$
\bea
E_0^\infty={1\over 4}(a-b)(a-c)+{1\over 2}\big[(a-c) +(c-b)(N+1)\big]\bfloor{\frac{c(p'-p)}{p'}},\quad(\sigma_0,\sigma_N,\sigma_{N+1})=(a,b,c)
\eea

\section{One-Point Functions of the Forrester-Baxter Models}\label{OPFs}

In this section we discuss the One Point Functions (OPFs) and Generalized Order Parameters (GOPs)
of the non-unitary Forrester-Baxter models~\cite{FB84} with $m'-m>1$. For these lattice models, some face weights are negative
and so there is no probabilistic interpretation of the Gibbs distribution. This is in contrast to the unitary Andrews-Baxter-Forrester models~\cite{ABF84}
with $m'-m=1$ for which the face weights are all positive, so there is a good probabilistic interpretation and the OPFs correspond to
Local Height Probabilities (LHPs). The non-unitary Forrester-Baxter models and the interpretation of their OPFs have been further discussed by various authors~\cite{non unitary}.

\subsection{Results of Forrester-Baxter}
The one point functions $\langle \delta(\sigma_0,s) \rangle_b$ of the Forrester-Baxter models are related to derivatives, with respect to
the conjugate  thermodynamic fields, of the logarithm of the free energy. In practice, however, it is better to calculate these quantities
as thermodynamic averages
\bea
P_{r,s}=\langle \delta(\sigma_0,s) \rangle_b=\lim_{N\to\infty}{\Tr S ABCD\over \Tr ABCD}\label{CTMform}
\eea
These quantities can be calculated~\cite{FB84} using Corner Transfer Matrices (CTMs)~\cite{BaxBook}.
The geometry for the CTM calculation is shown in Figure~2.
The corner transfer matrices are
\bea
A=A(u),\quad B=A(\lambda-u),\quad C=A(u),\quad D=A(\lambda-u)
\eea
and the center height is fixed to the value $s$ by the constant diagonal matrix $S$.
Since nearest neighbour heights differ
 by $\pm 1$, the parity of $N$ through which the limit is taken in (\ref{CTMform}) must be such that it is possible to get from $\sigma_0=s$ to $\sigma_N=b$ in $N$ steps with $\sigma_{N+1}=b+1$. By the $\mathbb{Z}_2$ sublattice symmetry, for groundstate OPFs with $b=b(r)$, it makes no difference if the other parity of $N$ is taken with $\sigma_N=b+1$ and $\sigma_{N+1}=b$.

Explicitly, the groundstate OPFs of the Forrester-Baxter models are given by (2.4.13) of \cite{FB84}
\bea
\begin{array}{rcl}
P_{r,s}(q)=\langle \delta(\sigma_0,s) \rangle_b&=&\disp
\frac{q^{(s-r)(s-r-1)/4}X_{r,s}(q)E(q^{s\over 2},q^{\frac{m'}{2(m'-m)}})}{\sum_{a}q^{(a-r)(a-r-1)/4}X_{r,a}(q)E(q^{a\over 2},q^{\frac{m'}{2(m'-m)}})}\\[16pt]
&=&\disp\frac{q^{(s-r)(s-r-1)/4}(q)_\infty\,X_{r,s}(q)E\big(q^{s/2},q^{\frac{m'}{2(m'-m)}}\big)}
{E(\!-q^{\frac{1}{2}},q^2)E\big(q^{r/2},q^{\frac{m}{2(m'-m)}}\big)},\qquad
b=b(r)
\end{array}
\label{FBOPFs}
 \eea
 where the (low-temperature) nome is $q=e^{-4\pi\lambda/\epsilon}$ and the sum in the denominator is restricted to $a=s$ mod 2.
 The second form follows from a denominator identity~\cite{FB84}.
 The thermodynamic limit $N\to\infty$ is taken through values with
 parity $N=a-b$ mod 2 so that
 \bea
\begin{array}{rcl}
X_{r,a}(q)\!\!&=&\!\!\disp{\lim_{N\to\infty} X^{(N)}_{ab\{b+1\}}(q)}= q^{\frac{c}{24}-\Delta^{m,m'}_{r,a}}\mbox{ch}^{m,m'}_{r,a}(q)\\[8pt]
\!\!&=&\!\!\disp\frac{1}{(q)_\infty}[E(\!-q^{mm'+rm'-am},q^{2mm'})-q^{ra}E(\!-q^{mm'+rm'+am},q^{2mm'})]
\end{array}\label{Xra}
\eea
where the elliptic function $E(x,q)$ is given in Appendix~B.
The low-temperature limit is given by
\bea
\lim_{q\to 0}P_{r,s}(q)=\lim_{q\to 0}\langle \delta(\sigma_0,s) \rangle_b=\delta(r,s)+\delta(r+1,s),\qquad b=b(r)
\eea
Alternatively, the OPFs can be expressed in terms of the original nome $t=e^{-\epsilon}$ by taking a conjugate modulus transformation.
Explicitly, from Forrester-Baxter~\cite{FB84} (3.1.18c)
\bea
P_{r,s}(t)=\frac{\vartheta_1(s\lambda,t)\big[\vartheta_3(\frac{\pi}{2}(\frac{r}{m}-\frac{s}{m'}),
t^{\frac{1}{4m(m'-m)}})-\vartheta_3(\frac{\pi}{2}(\frac{r}{m}+\frac{s}{m'}),
t^{\frac{1}{4m(m'-m)}})\big]}{m'\vartheta_4(0,t^{\frac{m'}{m'-m}})\vartheta_1(\frac{\pi
r(m'-m)}{m},t^{\frac{m'}{m}})} \label{prob}
\eea
where the elliptic functions $\vartheta_3$ and $\vartheta_4$ are given in Appendix~B.

There are $\frac{1}{2}(m-1)(m'-1)$ independent OPFs
$P_{r,s}$ because they satisfy the Kac table symmetry $P_{m-r,m'-s}=P_{r,s}$.
Here it is convenient to restrict $(r,s)$ to the fundamental domain
\bea
\mathcal{J}=\{(r,s)\,|\,
1\leqslant r\leqslant m-1, 1 \leqslant s\leqslant m'-1,m'r+ms
\leqslant mm' \} \label{set}
\eea
By Bezout's lemma, there always exists a unique index pair $(r_0, s_0)\in \mathcal{J}$ such that
\bea
\Delta^{m,m'}_{r_0,s_0} =\frac{1-(m'-m)^2}{4mm'}=\mathop{\mbox{Min}}_{(r,s)\in{\cal J}} \Delta^{m,m'}_{r,s}
\eea
which is the minimum conformal dimension in the Kac table of the minimal model.
Following Forrester-Baxter, we note that the non-unitary OPFs (\ref{prob}) all diverge at criticality $t\to 0$
\bea
P_{r,s}(t) \sim
(t^{\frac{\pi}{\lambda}})^{\Delta_{r_0,s_0}^{m,m'}},\qquad \Delta_{r_0,s_0}^{m,m'}={\frac{1-(m'-m)^2}{4mm'}}<0,\qquad m'-m>1
\eea
As pointed out by Forrester-Baxter, this divergence cannot happen for a probabilistic distribution but can occur for the
non-unitary minimal models because these models are unphysical in the sense that some Boltzmann weights are negative.
By contrast, for the unitary cases with $m'-m=1$, the $P_{r,s}$
are (normalized) probabilities bounded between 0 and 1. Forrester-Baxter remark that,
since the $P_{r,s}$ diverge at criticality, it is not possible to
define exponents of the order parameters in the usual sense.
Nonetheless, we will define new observables simply related to
Generalized Order Parameters (GOPs) which will exhibit well-defined
critical exponents.


\subsection{Generalized order parameters and critical exponents}

To define GOPs, we will make use of the modular properties of the Virasoro characters.
The Forrester-Baxter OPFs can be rewritten as
\bea
 P_{r,s}(q)=\frac{q^{\frac{c}{24}-\Delta^{m,m'}_{r,s}+\frac{(s-r)(s-r-1)}{4}}
E(q^{\frac{s}{2}},q^{\frac{m'}{2(m'-m)}})(q)_\infty}{E(-q^{\frac{1}{2}},q^2)E(q^{\frac{r}{2}},q^{\frac{m}{2(m'-m)}})}\
\mbox{ch}^{m,m'}_{r,s}(q) \label{simpPrs}
\eea
To carry out the conjugate modulus
transformation, we can use the conjugate modulus transformations of
Appendix~B and the modular matrix $\mathcal{S}$ of the minimal models~\cite{FMS1987}
\bea
\mathcal{S}_{rs;r's'}=
\sqrt{\frac{8}{mm'}}\,(-1)^{(r'+s')(r+s)}
\sin\frac{\pi(m'-m)rr'}{m}\sin\frac{\pi(m'-m)ss'}{m'}\label{Smatrix}
\eea
to obtain
\begin{align}
P_{r,s}(t)&=
\sqrt{\dfrac{2m}{m'}}\dfrac{\eta(t^\frac{m'}{m'-m})\vartheta_1(\frac{s\pi(m'-m)}{m'},t)}{\vartheta_4(0,t^{\frac{m'}{m'-m}})\vartheta_1(\frac{\pi
r(m'-m)}{m},t^{\frac{m'}{m}})} \sum_{(r's')\in \mathcal{J}}\mathcal{S}_{rs;r's'}\,
\mbox{ch}_{r',s'}^{m,m'}(t^{\frac{m'}{m'-m}})\label{critPrs}
\end{align}
where the Dedekind eta function is $\eta(q)=q^{\frac{1}{24}}(q)_\infty$.
The modular matrix $\mathcal{S}$ is real orthogonal and
has the conjugation matrix ${\cal C}=\mathcal{S}^2=I$.

Generalizing Huse~\cite{Huse}, we introduce Generalized Order
Parameters (GOPs) as linear combinations of the OPFs
\bea
R_{r'',s''}=\sum_{(r,s)\in \mathcal{J}} \mathcal{S}_{r''s'';rs}\;
\frac{\sin\frac{\pi r(m'-m)}{m}}{\sin\frac{\pi s(m'-m)}{m'}}\;P_{r,s} \label{GOPs}
\eea
The modular matrix
coefficients are introduced to effectively counteract the
coefficients resulting from the modular matrix transformation to the
critical nome $t$ in (\ref{critPrs}), at least to leading orders. In substituting
(\ref{critPrs}) into (\ref{GOPs}) we observe that the $(r,s)$-dependent elliptic prefactors can be rewritten as
 \bea
\frac{\eta(t^\frac{m'}{m'-m})}{\vartheta_4(0,t^{\frac{m'}{m'-m}})}\frac{\vartheta_1(\frac{\pi
s(m'-m)}{m'},t)}{\vartheta_1(\frac{\pi r(m'-m)}{m},t^{\frac{m'}{m}})}
=(t^{\frac{\pi}{\lambda}})^{c\over 24}\,
\frac{\sin\frac{\pi s(m'-m)}{m'}}{\sin\frac{\pi r(m'-m)}{m}}\,
\frac{\hat{\vartheta}_1(\frac{\pi
s(m'-m)}{m'},t)}{\hat{\vartheta}_1(\frac{\pi
r(m'-m)}{m},t^{\frac{m'}{m}})}\,\frac{(t^2)_\infty(t^{\frac{2m'}{m'-m}})_\infty}{(t^{\frac{2m'}{m}})_\infty(t^\frac{m'}{m'-m})_\infty}
\eea
where
\bea
\hat{\vartheta}_1(u,t)=\frac{\vartheta_1(u,t)}{2
t^{\frac{1}{4}}(t^2)_\infty \sin u}=\prod_{n=1}^\infty
(1-2t^{2n}\cos 2u+t^{4n})
\eea
and we have used the relation
$\vartheta_4(0,t)=(t)_\infty(t)_\infty/(t^2)_\infty$. It is now
apparent that the ratio of trigonometric functions cancels out of
the expression for the GOPs $R_{r'',s''}$. As a consequence of  the
involution property $\mathcal{S}^2=I$, the coefficients of the
modular matrix combine in such a way as to leave a contribution from a single
character, up to the correction term arising from
\bea
\frac{\hat{\vartheta}_1(\frac{\pi
s(m'-m)}{m'},t)}{\hat{\vartheta}_1(\frac{\pi
r(m'-m)}{m},t^{\frac{m'}{m}})}\sim 1+O(t^2)
\eea

Clearly, $P_{r,s}$ and $R_{r,s}$ have convergent Taylor expansions about $t=0$ in the variable $t^{1\over m(m'-m)}$. Defining new
observables ${\cal O}_{r,s}$ as ratios of the GOPs yields the Taylor expansions
\bea
{\cal O}_{r,s}=\frac{R_{r,s}}{R_{1,1}} \sim
(t^\frac{\pi}{\lambda})^{\Delta_{r,s}^{m,m'}}+(t^\frac{\pi}{\lambda})^{\Delta_{r_0,s_0}^{m,m'}}O(t^2)\label{observables}
\eea
where $t^2$ measures the departure from criticality. The coefficient of the leading term is precisely 1 and the second term is the correction relative to $R_{1,1}$.
We conclude that the associated critical exponents are
\bea
{\cal O}_{r,s}\sim (t^2)^{\beta_{r,s}},\qquad \beta_{r,s}=(2-\alpha)\Delta_{r,s}^{m,m'}=\frac{(rm'-sm)^2-(m'-m)^2}{8m(m'-m)}
 \eea
where the free energy exponent is given~\cite{FB84} by $2-\alpha={\pi\over 2\lambda}=\frac{m'}{2(m'-m)}$ corresponding to the $\varphi_{1,3}$ perturbation off-criticality as in Appendix~A. A plot of the observables ${\cal O}_{r,s}$ is shown in Figure~4 for the minimal model ${\cal M}(4,7)$.
\begin{figure}[htb]
\centerline {
\includegraphics[width=3.2in]{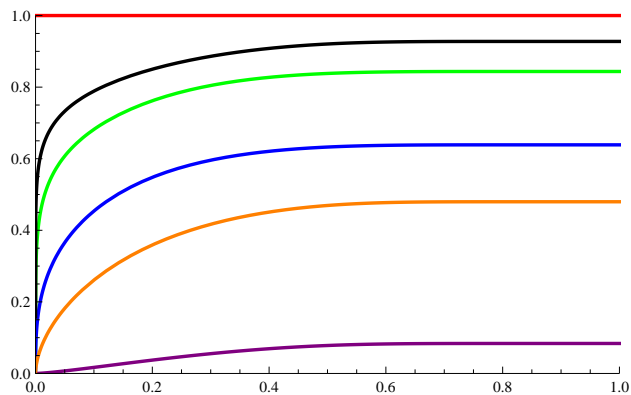}
}
\caption{Plot of the observables ${\cal O}_{r,s}$, as a function of $t$, for the minimal model ${\cal M}(4,7)$. Reading from top to bottom, we plot
${\cal O}_{1,1}, {1\over {\cal O}_{2,3}}, {1\over {\cal O}_{1,2}}, {\cal O}_{1,3}, {\cal O}_{2,2}, {\cal O}_{1,4}$ corresponding to
$|\Delta_{r,s}^{4,7}|=0, {5\over 112}, {1\over 14}, {1\over 7}, {27\over 112}, {9\over 14}$ in increasing order with critical exponents
$\beta_{r,s}={7\over 6}\,\Delta_{r,s}^{4,7}$. As expected for order parameters, these observables are nonegative, vanish at criticality and are increasing functions of $t$.}
\end{figure}

In this way, we have constructed
observables with associated critical exponents for all conformal
weights satisfying
 \bea
 \Delta_{r,s}^{m,m'}<\Delta_{r_0,s_0}^{m,m'}+\frac{2(m'-m)}{m'}={1+(m'-m)(9m-m')\over 4m m'}\label{observe}
\eea
These occur for $(r,s)$ in the Kac table satisfying
$(m'r-ms)^2<1+8m(m'-m)$. In fact, we have constructed
one such observable for each position $(r,s)$ in the Kac table but the conformal
weights not satisfying (\ref{observe}) are masked because the
correction terms are of lower order in the Taylor series expansion of (\ref{observables}).
Some Kac tables showing the relevant exponents are shown in Table~2.
\begin{table}[htbp]
\begin{center}
\psset{unit=.85cm}
{\vspace{.5cm}
{\small
\begin{pspicture}(0,0)(1,4)
\myframe(0,0)(1,4)
\psgrid[gridlabels=0pt,subgriddiv=1]
\rput(.5,3.5){$0$}
\rput(.5,2.5){$-\frac 15$}
\rput(.5,1.5){$-\frac 15$}
\rput(.5,.5){$0$}
{\color{blue}
\rput(.5,-.5){$1$}
\rput(1.5,-.5){$r$}
\rput(-.5,.5){$1$}
\rput(-.5,1.5){$2$}
\rput(-.5,2.5){$3$}
\rput(-.5,3.5){$4$}
\rput(-.5,4.5){$s$}}
\end{pspicture}
\qquad\qquad\qquad
\begin{pspicture}(0,0)(3,6)
\psframe[linewidth=0pt,fillstyle=solid,fillcolor=lightlightblue](0,0)(3,6)
\myframe(0,0)(1,4)
\myframe(1,1)(2,5)
\myframe(2,2)(3,6)
\psgrid[gridlabels=0pt,subgriddiv=1]
\rput(.5,5.5){$\frac{5}2$}\rput(1.5,5.5){$\frac {13}{16}$}\rput(2.5,5.5){$0$}
\rput(.5,4.5){$\frac{10}7$}\rput(1.5,4.5){$\frac{27}{112}$}\rput(2.5,4.5){$-\frac 1{14}$}
\rput(.5,3.5){$\frac{9}{14}$}\rput(1.5,3.5){$-\frac{5}{112}$}\rput(2.5,3.5){$\frac 17$}
\rput(.5,2.5){$\frac 17$}\rput(1.5,2.5){$-\frac{5}{112}$}\rput(2.5,2.5){$\frac{9}{14}$}
\rput(.5,1.5){$-\frac 1{14}$}\rput(1.5,1.5){$\frac {27}{112}$}\rput(2.5,1.5){$\frac{10}7$}
\rput(.5,.5){$0$}\rput(1.5,.5){$\frac{13}{16}$}\rput(2.5,.5){$\frac{5}{2}$}
{\color{blue}
\rput(.5,-.5){$1$}
\rput(1.5,-.5){$2$}
\rput(2.5,-.5){$3$}
\rput(3.5,-.5){$r$}
\rput(-.5,.5){$1$}
\rput(-.5,1.5){$2$}
\rput(-.5,2.5){$3$}
\rput(-.5,3.5){$4$}
\rput(-.5,4.5){$5$}
\rput(-.5,5.5){$6$}
\rput(-.5,6.5){$s$}}
\end{pspicture}
\qquad\qquad\qquad
\begin{pspicture}(0,0)(4,6)
\psframe[linewidth=0pt,fillstyle=solid,fillcolor=lightlightblue](0,0)(4,6)
\myframe(0,0)(1,3)
\myframe(1,1)(2,4)
\myframe(2,2)(3,5)
\myframe(3,3)(4,6)
\psgrid[gridlabels=0pt,subgriddiv=1]
\rput(.5,5.5){$\frac{15}4$}\rput(1.5,5.5){$\frac {9}{5}$}\rput(2.5,5.5){$\frac{11}{20}$}\rput(3.5,5.5){$0$}
\rput(.5,4.5){$\frac{16}7$}\rput(1.5,4.5){$\frac{117}{140}$}\rput(2.5,4.5){$\frac 3{35}$}\rput(3.5,4.5){$\frac 1{28}$}
\rput(.5,3.5){$\frac{33}{28}$}\rput(1.5,3.5){$\frac{8}{35}$}\rput(2.5,3.5){$-\frac {3}{140}$}\rput(3.5,3.5){$\frac{3}7$}
\rput(.5,2.5){$\frac 37$}\rput(1.5,2.5){$-\frac{3}{140}$}\rput(2.5,2.5){$\frac{8}{35}$}\rput(3.5,2.5){$\frac{33}{28}$}
\rput(.5,1.5){$\frac 1{28}$}\rput(1.5,1.5){$\frac {3}{35}$}\rput(2.5,1.5){$\frac{117}{140}$}\rput(3.5,1.5){$\frac{16}7$}
\rput(.5,.5){$0$}\rput(1.5,.5){$\frac{11}{20}$}\rput(2.5,.5){$\frac{9}{5}$}\rput(3.5,.5){$\frac{15}4$}
{\color{blue}
\rput(.5,-.5){$1$}
\rput(1.5,-.5){$2$}
\rput(2.5,-.5){$3$}
\rput(3.5,-.5){$4$}
\rput(4.5,-.5){$r$}
\rput(-.5,.5){$1$}
\rput(-.5,1.5){$2$}
\rput(-.5,2.5){$3$}
\rput(-.5,3.5){$4$}
\rput(-.5,4.5){$5$}
\rput(-.5,5.5){$6$}
\rput(-.5,6.5){$s$}}
\end{pspicture}
}}
\end{center}
\caption{Kac tables of conformal weights $\Delta_{r,s}^{m,m'}$ for the minimal models ${\cal M}(2,5)$, ${\cal M}(4,7)$, ${\cal M}(5,7)$ respectively.
The boxes of the conformal dimensions \mbox{$\Delta_{r,s}< {1+(m'-m)(9m-m')\over 4m m'}$} that are accessible, using Corner Transfer Matrices to calculate off-critical observables, are shaded in pink and occur in the band where $(r,s)$ satisfy $(m' r-ms)^2< 1+8m(m'-m)$.}
\end{table}

\subsection{One point functions of the logarithmic minimal models}

In this section, we apply the logarithmic limit (\ref{offlogLimit}) to the OPFs and GOPs of the Forrester-Baxter models.
This is straightforward in the low-temperature nome. From (\ref{simpPrs}), we find the limiting OPFs
\bea
 P_{r,s}^\infty(q)=\frac{q^{\frac{c}{24}-\Delta^{p,p'}_{r,s}+\frac{(s-r)(s-r-1)}{4}}
E(q^{\frac{s}{2}},q^{\frac{m'}{2(p'-p)}})(q)_\infty}{E(-q^{\frac{1}{2}},q^2)E(q^{\frac{r}{2}},q^{\frac{p}{2(p'-p)}})}\
\chi^{p,p'}_{r,s}(q),\qquad r,s=1,2,3,\ldots \label{Prs}
\eea
The difficulty is in implementing a conjugate modulus transformation to the critical nome $t$
which is necessary to obtain the critical exponents. The problem derives from the fact that there is no simple
conjugate modulus transformation on the infinity of the quasi-rational Kac characters $\chi_{r,s}^{p,p'}(q)$.
Indeed, it is easily seen that all of the entries (\ref{Smatrix}) of the ${\cal S}$ matrix have a common prefactor $\sqrt{8/m m'}$ which vanishes in the logarithmic limit.
As a consequence, the conjugate modulus form (\ref{critPrs}) of the OPFs and the GOPs (\ref{GOPs}) (which both involve infinite sums)
are not well behaved in the logarithmic limit. Nevertheless, the logarithmic limit of the observables  ${\cal O}_{r,s}$ (\ref{observables})
(in which the problematic prefactors cancel out in the ratio) are well defined and admit a Taylor expansion about $t=0$\\
\bea
{\cal O}_{r,s}^\infty=\lim_{m,m'\to\infty, \  {m\over m'}\to {p\over p'}} \frac{R_{r,s}}{R_{1,1}} \sim
(t^\frac{\pi}{\lambda})^{\Delta_{r,s}^{p,p'}}+(t^\frac{\pi}{\lambda})^{\Delta_{r_0,s_0}^{p,p'}}O(t^2)\label{logobservables},\qquad \lambda={(p'-p)\pi\over p'}
\eea
This is just the logarithmic limit of the Taylor expansion (\ref{observables}) for the Forrester-Baxter observables.
The limits of these Taylor expansions were checked using Mathematica~\cite{Wolfram}.

We conclude that the associated logarithmic critical exponents are
\bea
{\cal O}_{r,s}^\infty \sim (t^2)^{\beta_{r,s}},\qquad \beta_{r,s}=(2-\alpha)\Delta_{r,s}^{p,p'}=\frac{(rp'-sp)^2-(p'-p)^2}{8p(p'-p)}
 \eea
where the free energy exponent of the off-critical logarithmic minimal models
$2-\alpha={\pi\over 2\lambda}=\frac{p'}{2(p'-p)}$, corresponding to the $\varphi_{1,3}$ perturbation off-criticality, is obtained in Appendix~A. In this way, we have constructed
limiting observables with associated critical exponents for all conformal
weights satisfying
 \bea
 \Delta_{r,s}^{p,p'}<\Delta_{r_0,s_0}^{p,p'}+\frac{2(p'-p)}{p'}={(p'-p)(9p-p')\over 4p p'}\label{logobserve}
\eea
These occur for $(r,s)$ in the infinitely extended Kac tables satisfying
$(p'r-ps)^2<8p(p'-p)$.
The Kac tables of critical dense polymers ${\cal LM}(1,2)$ and critical percolation ${\cal LM}(2,3)$
showing the relevant exponents are presented in Table~1.

\section{Conclusion}

The face weights of the Forrester-Baxter models (associated with the non-unitary minimal models) can be decomposed by incorporating internal degrees of freedom, in the form of loop segments or bonds, that are suitable to describe generalized percolation or cluster properties. In this paper, we have used Corner Transfer Matrices to argue that the logarithmic limit (\ref{offlogLimit}) of the non-unitary minimal models ${\cal M}(m,m';t)$ provides the integrable $\varphi_{1,3}$ off-critical perturbation of the logarithmic minimal models ${\cal LM}(p,p')$.
Generalized models of polymers and percolation can thus be solved exactly off-criticality.
To support this assertion we have shown that, in accordance with the Kyoto correspondence principle~\cite{Kyoto}, the limiting one-dimensional configurational sums yield the finitized quasi-rational Kac characters of the logarithmic theories.
In addition, we have identified off-critical observables (\ref{observables}) in the minimal models that, in the logarithmic limit, exhibit critical exponents associated  with conformal weights $\Delta_{r,s}^{p,p'}$ in the infinitely extended Kac table of the logarithmic theory for all Kac labels $(r,s)$ satisfying
$(ps-p'r)^2<8p(p'-p)$.
We emphasize again that, although we call these limiting models off-critical {\it logarithmic\/} minimal models, these off-critical models are not expected to exhibit the  indecomposable structures associated with critical logarithmic CFTs.

We believe that the evidence supporting our assertion is strong even though it is in a sense indirect.
Perhaps our results could be obtained more directly by considering the unrestricted SOS models~\cite{BaxBook} with generic crossing parameter $\lambda$.
If $\lambda$ is taken to be a rational fraction of $\pi$, this lattice model truncates to an RSOS model. However, we expect that the same off-critical logarithmic minimal models can be obtained if the logarithmic limit to a crossing parameter which is a rational fraction of $\pi$ is taken (through irrational fractions of $\pi$) after the thermodynamic limit.

The prototypical example of the logarithmic minimal models is percolation ${\cal LM}(2,3;t)$. The traditional approach to percolation~\cite{Stauffer} is to take either the limit $n\to 0$~\cite{nVec} of the $n$-vector models or the limit $Q\to 1$~\cite{FKclusters} of the $Q$-state Potts models. The drawback of these approaches is that they require an analytic continuation to non-integer $n$ or $Q$. A benefit of the current approach is that the logarithmic limit is completely under control since it is taken through a sequence of well-defined integrable lattice models.

The universal amplitude ratio $f_{\text{sing}}/M^2$ for the $\varphi_{1,3}$ perturbation of the minimal models ${\cal M}(m,m')$ has been obtained from field theory by Lukyanov and Zamolodchikov~\cite{LZ}. Here the mass $M$ is  proportional to the inverse correlation length. Universal amplitude ratios~\cite{DelfViti2010}, for either the $\varphi_{1,3}$, $\varphi_{2,1}$ or $\varphi_{1,2}$ perturbations of the minimal models, can also be obtained from the lattice following~\cite{Seaton}. It is then possible to obtain the universal amplitude ratios for the $\varphi_{1,3}$ perturbation of logarithmic minimal models, such as percolation ${\cal LM}(2,3)$. This is properly achieved by taking the logarithmic limit ($m\to\infty, m'\to\infty, m/m'\to p/p'$). Previously, this step would have been taken by formally replacing $(m,m')$ with $(p,p')$ without justification.
Using the logarithmic limit gives the same results but avoids the problem that, for example in the case $(p,p')=(2,3)$, the naive replacement ${\cal M}(2,3)$ corresponds to a trivial theory rather than the correct percolation theory ${\cal LM}(2,3)$.

\begin{figure}[tbp]
\psset{unit=.7cm}
\begin{center}
\begin{pspicture}(0,.3)(8,4.5)
\psline[linewidth=.25pt](.5,0)(7.5,0)
\psarc[linewidth=1pt](1,0){.5}{0}{180}
\psarc[linewidth=1pt](4,0){.5}{0}{180}
\psarc[linewidth=1pt](6,0){.5}{0}{180}
\psarc[linewidth=1pt](5,0){2.5}{0}{180}
\rput(0,-.4){\small $0$}
\rput(1,-.4){\small $1$}
\rput(2,-.4){\small $2$}
\rput(3,-.4){\small $3$}
\rput(4,-.4){\small $4$}
\rput(5,-.4){\small $5$}
\rput(6,-.4){\small $6$}
\rput(7,-.4){\small $7$}
\rput(8,-.4){\small $8$}
\end{pspicture}\qquad\qquad
\begin{pspicture}(0,.3)(8,4.5)
\psframe[linewidth=0pt,fillstyle=solid,fillcolor=lightlightblue](0,3)(8,4)
\psframe[linewidth=0pt,fillstyle=solid,fillcolor=lightlightblue](0,2)(8,3)
\psframe[linewidth=0pt,fillstyle=solid,fillcolor=lightlightblue](0,1)(8,2)
\psframe[linewidth=0pt,fillstyle=solid,fillcolor=lightlightblue](0,0)(8,1)
\multirput(1,0)(1,0){8}{\psline[linewidth=.5pt](0,0)(0,4)}
\psline[linewidth=1pt](0,0)(1,1)(2,0)(3,1)(4,2)(5,1)(6,2)(7,1)(8,0)
\rput(0,-.4){\small $0$}
\rput(1,-.4){\small $1$}
\rput(2,-.4){\small $2$}
\rput(3,-.4){\small $3$}
\rput(4,-.4){\small $4$}
\rput(5,-.4){\small $5$}
\rput(6,-.4){\small $6$}
\rput(7,-.4){\small $7$}
\rput(8,-.4){\small $8$}
\rput(-.3,0){\small $1$}
\rput(-.3,1){\small $2$}
\rput(-.3,2){\small $3$}
\rput(-.3,3){\small $4$}
\rput(-.3,4.2){\small $\vdots$}
\end{pspicture}
\end{center}
\caption{A link state of the logarithmic minimal model ${\cal LM}(p,p')$ in the $(r,s)=(1,1)$ vacuum sector and the corresponding Dyck path $\sigma=\{1,2,1,2,3,2,3,2,1\}$ of length $N=8$ on (one-sided) $A_\infty$  with the boundary conditions $(\sigma_0,\sigma_N)=(1,1)$. The height $\sigma_j$ of the path at position $j=0,1,\ldots,N$ is equal to 1 plus the number of loops segments directly above position $j$ in the link state. The total number of such link states or Dyck paths under this bijection for $N$ even is given by the Catalan numbers $C_N={N!\over (N/2)!(N/2+1)!}$.}
\end{figure}
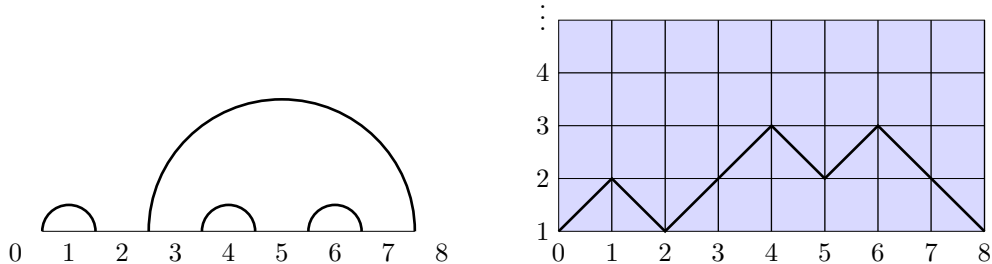

This paper opens up several directions for future research. First, rather than work with CTMs, it would be informative to apply the logarithmic limit (\ref{offlogLimit}) to the double-row transfer matrices of the minimal models on a strip both at criticality and off-criticality. The $(r,s)$ integrable boundary conditions for the non-unitary minimal models have not been studied in depth. Nevertheless at criticality, in the vacuum $(r,s)=(1,1)$ sector in which the boundary heights alternate between 1 and 2 on the left and right boundary edges of the strip, it is easy to see that the logarithmic limit of the conformal partition function is
\bea
Z^{m,m'}_{(1,1)|(1,1)}(q)=\mbox{ch}^{m,m'}_{r,s}(q)\to \chi_{r,s}^{p,p'}(q)=q^{-c/24+\Delta_{r,s}^{p,p'}}{(1-q)\over (q)_\infty}=Z^{p,p'}_{(1,1)|(1,1)}(q)
\eea
We observe that, for these boundary conditions, the finite-size double-row transfer matrices have the same dimension in the sense that there is a bijection between the unrestricted SOS or Dyck paths on (one-sided) $A_\infty$ and the link states of the critical logarithmic minimal models as shown in Figure~5. It is therefore natural to expect that the logarithmic limit maps the finitized minimal Virasoro characters (\ref{finvir}) onto the finitized quasi-rational Kac characters (\ref{logfinchar}). More explicitly, for the non-unitary minimal models with Yang-Baxter integrable $(r,s)$ boundary conditions, we expect
\bea
Z^{m,m',(N)}_{(1,1)|(r,s)}(q)=\mbox{ch}^{(N)}_{r,s}(q)\to\chi^{(N)}_{r,s}(q)=Z^{p,p',(N)}_{(1,1)|(r,s)}(q)
\eea
where the characters implicitly depend on $b=b(r)$ given by (\ref{b(r)}) and the dependence of these characters on $m,m'$ and $p,p'$ is suppressed.

Likewise, it would be of interest to study the logarithmic limit of the periodic row transfer matrices of the minimal models both at criticality and off-criticality. At criticality, this could shed some light on the conjectured torus modular invariant partition functions~\cite{GabRunkW,PZcoset} of the logarithmic minimal models.
The study of the logarithmic limit of the transfer matrices of the non-unitary models also opens up the possibility to obtain functional equations in the form of $T$- and $Y$-systems~\cite{TYsys}. In particular, the off-critical functional relations could give a means to derive massive Thermodynamic Bethe Ansatz~\cite{Zam} equations for the logarithmic minimal models. Lastly, it would be of interest to study the logarithmic limit of the $\varphi_{2,1}$ and $\varphi_{1,2}$ off-critical integrable perturbations of the minimal models. This would involve further study of the non-unitary dilute $A$ lattice models~\cite{diluteA}. We hope to come back to some of these problems later.

\appendix

\section{Logarithmic Limit of the Forrester-Baxter Free Energy}

The inversion relation method~\cite{Baxter82} has been used~\cite{FB84} to find the bulk partition function per site $\kappa(u)$, or equivalently, the free energy $f(u)=-\log\kappa(u)$ of the Forrester-Baxter models. In the notation of this paper,
\bea
\log \kappa(u)=u(\lambda-u)/\epsilon +2\sum_{n=1}^\infty
\frac{\cosh((2\lambda-\pi)\pi n /\epsilon)\sinh(n \pi u/
\epsilon)\sinh((\lambda-u)\pi n/ \epsilon)}{n \sinh(\pi^2
n/\epsilon)\cosh(\lambda \pi n/\epsilon)}
\eea
Using the Poisson summation formula, it can be shown that when $m'-m
\neq 1$ and $m'$ is {\it odd},
\begin{multline}
\log \kappa(u) =c(u, \lambda) -4\sum_{n=1}^\infty
\frac{t^{2n}}{n(1-t^{2n})}\frac{\sin(nu)\sin((\lambda-u)n)\cos(2\lambda
n)}{ \cos(\lambda n)}
\\-4
\sum_{n=1}^\infty
\frac{t^{(2n-1)\pi/\lambda}}{(2n-1)(1-t^{(2n-1)\pi/\lambda})}\
\sin\frac{(2n-1)\pi u}{\lambda}\cot\frac{(2n-1)\pi^2}{2\lambda}\label{bulk}
\end{multline}
where $c(u, \lambda)$ is independent of the thermal deviation
$t=e^{-\epsilon}$.
This also gives the correct leading-order singularity and amplitude
when $m'-m \neq 1$ and $m'$ is {\it even}, but the higher order terms in
equation (\ref{bulk}) are modified.
At the isotropic point $u=\frac{\lambda}{2}$, the singular part $f_{\text{sing}}(u)$ of the free energy is
\begin{equation}
f_{\text{sing}}(\tfrac{\lambda}{2}) \sim 4\,t^{\pi/ \lambda} \cot\frac{\pi^2}{2 \lambda}\label{sing}
\end{equation}
with the critical exponent as given in \cite{FB84}.

This confirms that the deviation from criticality $t^2$ corresponds to the field
$\varphi_{1,3}$ since, using the scaling relation on the associated
critical exponent,
\begin{equation}
2-\alpha=\frac{m'}{2(m'-m)}, \qquad
\Delta_t=\frac{1-\alpha}{2-\alpha}=\frac{2m-m'}{m'}=\Delta_{1,3}^{m,m'}\label{scale}
\end{equation}
Since the amplitude and exponent depend on $m, m'$ only through the
ratio $\lambda$, these results (\ref{sing}) and (\ref{scale}) also apply in the logarithmic limit with $\lambda={(p'-p)\pi\over p'}$
and
\bea
\Delta_t=\frac{2p-p'}{p'}=\Delta_{1,3}^{p,p'}\label{logscale}
\eea

The amplitude of the leading singular term of the free energy
\bea
\mathcal{A}_f=4 \cot\frac{\pi^2}{2 \lambda}=-4\tan\frac{\pi m}{2 (m'-m)}\label{amp}
\eea
is used in the determination of the universal amplitude ratio
$R_\xi^+ $, or its equivalent in perturbed conformal field theory
$f_{\text{sing}}/M^2$.  The mass $M$ is proportional to the inverse correlation length,
but it appears the correlation length $\xi$ of the Forrester-Baxter models has not yet been calculated. Nevertheless, (\ref{amp}) is consistent (at least in its
dependence on $m$ and $m'$) with the expression given by (14) of \cite{LZ}
for the $\varphi_{1,3}$ perturbation of $\mathcal{M}(m,m')$
\bea
{f_{\text{sing}}\over M^2}=-\frac{1}{4} \tan \frac{\pi m}{2 (m'-m)}
\eea

\section{Elliptic Functions}

For convenience we summarize the definitions and properties of the elliptic functions used throughout this paper.
The standard elliptic theta functions~\cite{GR} are
\begin{align}
\vartheta_1(u,t)&=2t^{1/4}\sin u
\prod_{n=1}^\infty (1-2t^{2n}\cos 2u+t^{4n})(1-t^{2n})\\
\vartheta_4(u,t)&=\prod_{n=1}^\infty (1-2t^{2n-1}\cos
2u+t^{4n-2})(1-t^{2n})
\end{align}
with $\vartheta_1(u, t)=\vartheta_2(\tfrac{\pi}{2}-u, t)$ and $\vartheta_3(u, t)=\vartheta_4(\tfrac{\pi}{2}-u, t)$. We make use of the following identity (15.4.26) of \cite{BaxBook} in writing the two forms of (\ref{Boltzmann3})
\bea
\vartheta_4(u)\vartheta_4(a\!-\!u)\vartheta_1(v)\vartheta_1(a\!-\!v)-\vartheta_4(v)\vartheta_4(a\!-\!v)\vartheta_1(u)\vartheta_1(a\!-\!u)=\vartheta_4(0)\vartheta_4(a)\vartheta_1(v\!-\!u)\vartheta_1(a\!-\!u\!-\!v)
\eea
The conjugate modulus transformations of the elliptic $\vartheta$-functions  are
\begin{align}
\vartheta_1(u,e^{-\epsilon})&=\sqrt\frac{\pi}{\epsilon}\, e^{-(u-\pi/2)^2/\epsilon}E(e^{-2 \pi u/\epsilon}, e^{-2\pi^2/\epsilon} )\\
\vartheta_4(u,e^{-\epsilon})&=\sqrt\frac{\pi}{\epsilon}\,e^{-(u-\pi/2)^2/\epsilon}E(-e^{-2 \pi u/\epsilon}, e^{-2\pi^2/\epsilon} )
\end{align}
where
\bea
E(x,q)=\sum_{k=-\infty}^{\infty} (-1)^kq^{k(k-1)/2}x^k=\prod_{n=1}^\infty(1-q^{n-1}x)(1-q^nx^{-1})(1-q^n)
\eea
The elliptic $\vartheta$ functions also have infinite sum representations but we do not use them in this paper.

\section*{Acknowledgements} This research is supported by the Australian Research Council. PAP thanks Laszlo Palla, Gabor Takacs and Zoli Bajnok of E\"otv\"os University, Budapest for hospitality and their interest in this problem. He also thanks Giuseppe Mussardo, Aldo Delfino and Jacopo Viti of SISSA, Trieste for hospitality and discussions.  Lastly, he thanks the Asia Pacific Center for Theoretical Physics, Pohang for hospitality.
KAS thanks the Department of Mathematics and Statistics of the University of Melbourne and the Australian Mathematical Sciences Institute (AMSI) for hosting her sabbatical leave, and La Trobe University for granting that leave.  We thank Timothy Trott for working through the details of the CTM recursions (\ref{logRecursion}) for the logarithmic minimal models. Lastly, we thank J{\o}rgen Rasmussen for a critical reading of the paper.

\end{document}